\documentclass[hyper,letterpaper,12pt]{article}
\usepackage{a4wide}

\usepackage{amsmath}
\usepackage{color}
\usepackage{graphicx}
\usepackage{amsfonts}
\usepackage{amssymb}
\usepackage{epstopdf}
\usepackage[
       colorlinks=true,
      filecolor=black,
      linkcolor=blue,
      citecolor=red,
      urlcolor=blue,
       linktocpage=true
            ]{hyperref}

\def\I{I }
\def\nn{{\nonumber}}
\def\be{\begin{equation}}
\def\ee{\end{equation}}
\def\beq{\begin{eqnarray}}
\def\eeq{\end{eqnarray}}
\def\ba{\beq\begin{array}{l}}
\def\ea{\end{array}\eeq}

\def\ep{\epsilon}

\def\p{\partial}
\def\s{\sigma}
\def\eps{\epsilon}

\def\sc{$\Sigma_c$}

\def\p{\partial}
\def\l{{\lambda}}
\def\a{{\alpha}}
\def\b{{\beta}}
\def\g{{ \gamma}}
\def\de{{\delta}}
\def\t{{ \theta}}
\def\e{{\epsilon}}

\def\x{\hat x}
\def\P{\hat P}
\def\r{\hat r}
\def\t{\hat t}
\def\v{\hat v}
\def\hT{{\hat{T}}}
\def\hp{\hat{\partial}}
\def\hpartial{\hat{\partial}}
\def\htau{{\hat{\tau}}}
\def\hs{\hat\sigma}

\def\homega{\hat\omega}

\def\hsigma{\hat\sigma}
\def\heta{\hat\eta}
\def\hc{\hat c}
\def\hF{\hat F}
\def\htheta{\hat \theta}
\def\hf{\hat f}
\def\erho{\mathrm{e}}
\def\hrho{\hat {\mathrm{e}}}

\def\d{\mathrm{d}}
\def\pp{\mathrm{p}}

\def\Pe{\mathrm{P}}
\def\Ha{\mathrm{H}}

\def\hPe{\hat{\mathrm{P}}}
\def\hHa{\hat{\mathrm{H}}}

\def\D{\partial}
\renewcommand{\g}{\gamma}

\renewcommand{\t}{\tau}
\renewcommand{\ep}{\epsilon}
\def\K{\mathcal{K}}

\renewcommand{\nn}{\nonumber}
\renewcommand{\be}{\begin{equation}}
\renewcommand{\ee}{\end{equation}}
\newcommand{\bea}{\begin{eqnarray}}
\newcommand{\eea}{\end{eqnarray}}
\renewcommand{\(}{\left(}
\renewcommand{\)}{\right)}
\renewcommand{\[}{\left[}
\renewcommand{\]}{\right]}

\begin{document}

\title{\bf \Large Petrov type \I  Condition and Dual Fluid Dynamics}

\author{\large
Rong-Gen Cai\footnote{E-mail: cairg@itp.ac.cn},
~~Li Li\footnote{E-mail: liliphy@itp.ac.cn},
~~Qing Yang\footnote{E-mail: yangqing@itp.ac.cn},
~~Yun-Long Zhang\footnote{E-mail: zhangyl@itp.ac.cn}\\
\\
\small State Key Laboratory of Theoretical Physics,\\
\small Institute of Theoretical Physics, Chinese Academy of Sciences,\\
\small %P.O. Box 2735,
Beijing 100190, People's Republic of China.\\}
%\date{\small March 25, 2013}
\date{\small February 11, 2014}
%\date{\small\today}

\maketitle

\begin{abstract}
Recently Lysov and Strominger [arXiv:1104.5502] showed that imposing
Petrov type \I  condition on a $(p+1)$-dimensional timelike
hypersurface embedded in a $(p+2)$-dimensional vacuum Einstein
gravity reduces the degrees of freedom in the extrinsic curvature of
the hypersurface to that of a fluid on the hypersurface, and that
the leading-order Einstein constraint equations in terms of the mean
curvature of the embedding give the  incompressible Navier-Stokes
equations of the dual fluid. In this paper we show that the
non-relativistic fluid dual to vacuum Einstein gravity does not
satisfy the Petrov type \I  condition at next order, unless
additional constraint such as the irrotational condition is added.
In addition, we show that this procedure can be inversed to derive
the non-relativistic hydrodynamics with higher order corrections
through imposing the Petrov type \I  condition, and that some second
order transport coefficients can be extracted, but the dual ``Petrov
type \I  fluid'' does not match the dual fluid constructed from the
geometry of vacuum Einstein gravity in the non-relativistic limit.
We discuss the procedure both on the finite cutoff surface via the
non-relativistic hydrodynamic expansion and on the highly
accelerated surface via the near horizon expansion.
\end{abstract}

\newpage

\tableofcontents

\section{Introduction}

In the non-relativistic hydrodynamic limit, a correspondence between
 the nonlinear solutions of the Einstein equations and
incompressible Navier-Stokes equations is constructed
in~\cite{Bredberg:2010ky,Bredberg:2011jq,Compere:2011dx} where an
intrinsically flat finite cutoff surface and regularity on the
future horizon are imposed. Two equivalent presentations of the
non-linear perturbed gravity solution and dual fluid expansion are
given, one is for the dual fluid living on a finite cutoff surface
via non-relativistic hydrodynamic expansion, the other is on the
highly accelerated surface via near horizon expansion. This relation
is further shown to be universal for the geometry with sphere
horizon~\cite {Bredberg:2011xw,Anninos:2011zn} and with higher
curvature
corrections~\cite{Cai:2011xv,Chirco:2011ex,Niu:2011gu,Cai:2012vr,Zou:2013ix}.
And the dual incompressible Navier-Stokes equations are found to be
corrected at leading order when a non-trivial gravitational
Chern-Simons term appears in the bulk~\cite{Cai:2012mg}. More
generally, the gravity is related with a fluid without gravity in
one lower dimension, and related works can also be found in
\cite{Kuperstein:2011fn, Brattan:2011my,Eling:2011ct,
Compere:2012mt,Eling:2012ni,Matsuo:2012pi,Bai:2012ci,Berkeley:2012kz,Caldarelli:2012hy},
which show their close relation with the fluid dynamics from
membrane
paradigm~\cite{Price:1986yy,Gourgoulhon:2005ng,Gourgoulhon:2005ch,Eling:2009pb,Eling:2009sj},
as well as the fluid/gravity correspondence from
holography~\cite{Policastro:2002se,Kovtun:2003wp,Bhattacharyya:2008jc,Bhattacharyya:2008kq,Hubeny:2011hd}.

It was noted in \cite{Bredberg:2011jq} that the nonlinear solution of vacuum Einstein gravity is of an
algebraically special Petrov type \cite{exac,Coley:2004jv,Coley:2007tp}, and the procedure was reversed
via the near horizon expansion in~\cite{Lysov:2011xx} to derive the dual hydrodynamics. The Petrov type
 \I  condition is imposed to reduce the Einstein equations to the incompressible Navier-Stokes equations
 in one lower dimension. The universal fixed-point behavior of the near-horizon scaling in general relativity
 is shown
  to be the same as that of hydrodynamic scaling in fluid dynamics~\cite{Lysov:2011xx}. This condition is
 expected to be equivalent to the regularity on the future horizon, and the framework has also been generalized
 to the
  highly accelerated surface which is spatially curved, and to the case with  cosmological constant
 and Maxwell field in the
bulk~\cite{Huang:2011he,Huang:2011kj,Zhang:2012uy}.

Note that in those works only the nontrivial leading order has been
considered, we are here going to generalize the procedure to higher
order to see whether the equivalence still holds or not. In the
frame which is associated with a hypersurface where the dual fluid
lived on, we find that the non-relativistic fluid dual to the
non-linear solution of vacuum Einstein gravity from boost
transformation does not satisfy the Petrov type \I  condition at the
next order, unless additional constraint is added such as the
irrotational condition.  We also inverse this procedure by imposing
the Petrov type \I  condition on the fluid stress tensor, and then
obtain the non-relativistic hydrodynamics with higher order
corrections. But we see that the dual ``Petrov type \I  fluid'' can
not match  the dual fluid of vacuum Einstein gravity constructed in
the non-relativistic limit. We study the procedure in two equivalent
expansions: one is the non-relativistic hydrodynamic expansion
associated with a finite cutoff surface, the other is the near
horizon expansion associated with a highly accelerated surface.

This paper is organised as follows. In section \ref{Sec2}, a simple review of the Petrov type \I
condition is given. In section \ref{Sec3}, the higher order non-relativistic stress tensor dual to
 vacuum Einstein gravity is used to check the Petrov type \I  condition. Then the logic is turned
 around and the Petrov type \I  condition is imposed to reduce the gravity to the dual non-relativistic
  hydrodynamics. In section \ref{Sec4}, an alternative presentation of this procedure in the near horizon
   expansion is discussed. The results and discussions are given in section \ref{Sec5}.

\section{Petrov type \I  condition}
\label{Sec2}

Firstly, we give a simple review of the Petrov type \I  condition
with respect to the ingoing and outgoing pair of null vectors whose
tangents to a timelike hypersurface generate time
translations~\cite{Lysov:2011xx}. Introducing the $(p+2)$
Newman-Penrose-like vector fields, \beq
\ell^2=k^2=0,\quad(k,\ell)=1,\quad(m_i,k)=(m_i,\ell)=0,\quad
(m_i,m_j)=\delta_{ij}, \eeq the spacetime is Petrov type
\I ~\cite{Coley:2004jv,Coley:2007tp} if for some choice of frame,
\be\label{petrovij} C_{(\ell)i(\ell)j}=0,\quad
C_{(\ell)i(\ell)j}\equiv \ell^\mu m_i^\nu \ell^\alpha m_j^\beta
C_{\mu\nu\alpha\beta}. \ee Consider a timelike $(p+1)$-dimensional
hypersurface \sc\ with flat intrinsic metric \be
ds_{p+1}^2=\gamma_{ab}dx^adx^b=
-(dx^0)^2+\delta_{ij}dx^idx^j,~~~~i,j=1,...,p, \ee and extrinsic
curvature $K_{ab}$. The hypersurface is embedded in a
$(p+2)$-dimensional vacuum Einstein spacetime that \be
G_{\mu\nu}=0,~~~\mu,\nu=0,...,p+1.\label{Einstein}\ee Choosing the
frame that \be\label{frame} m_i=\p_i,\;\; \sqrt{2}\ell = \p_0
-n,\;\; \sqrt{2}k=-\p_0-n, \ee where $n$ is the spacelike unit
normal to the hypersurface, and $\p_i,\p_0$ are the tangent vectors
to \sc~\cite{Lysov:2011xx}, one has \beq\label{typeI}
2C_{(\ell)i(\ell)j}= (K-K_{00})K_{ij}+2K_{0i}K_{0j} +2\p_0
K_{ij}-K_{ik}K_{~j}^k -\p_i K_{0j}-\p_j K_{0i}, \eeq where the
following  projections to \sc\ have been used
\begin{align} \label{swey} \g_a^\a \g_b^\b\g_c^\g \g_d^\delta C_{\a\b\g\de}&=K_{ad}K_{bc} -K_{ac}K_{bd},\nn\\
\g_a^\a \g_b^\b \g_c^\g n^\de C_{\a\b\g\de} &= \p_aK_{bc}-\p_bK_{ac},  \nn\\
\g_a^\a n^\b \g_c^\g n^\de C_{\a\b\g\de}&=K K_{ac}-K_{ab}K^b_{~c}\,,
\end{align}
with $\gamma^\alpha_a=\delta^\alpha_a-n_a n^\alpha$. The Petrov type \I  condition (\ref{petrovij}) imposes
$(p-1)(p+2)/2$ constraints on the $(p+1)(p+2)/2$ components of $K_{ab}$, or determines the trace-free part of
$K_{ij}$ in terms of $K,~K_{00}$ and $K_{0i}$. This leaves $(p+2)$ independent components, which are exactly
the number of components of a fluid with a local energy density, pressure and velocity.
The dual fluid is described by the Brown-York stress tensor on the hypersurface,
\be T_{ab}=2(K\gamma_{ab}-K_{ab}). \label{BYcut}\ee
The Hamiltonian constraint of vacuum Einstein equations
\beq\label{Ham} 2G_{\mu\nu}n^\mu n^\nu |_{\Sigma_c} =(K^2-K_{ab}K^{ab})=0\Longrightarrow
T^2-p\,T_{ab}T^{ab}=0, \eeq
can be viewed as the equation of state for the dual fluid relating the pressure and energy density.
On the other hand, the $(p+1)$ momentum constraint equations
\beq\label{Mon}2 G_{\mu b}n^\mu|_{\Sigma_c}=2(\p^aK_{ab}-\p_bK)=0\Longrightarrow
\p^aT_{ab}=0, \eeq
give us the equations of motion for the dual fluid.

\section{On finite cutoff surface}
%{Non-relativistic hydrodynamic expansion}
\label{Sec3}

In this section, with the non-relativistic stress tensor of fluid
dual to vacuum Einstein gravity at finite cutoff surface given
in~\cite{Compere:2011dx}, we will firstly check whether the Petrov
type \I  condition is satisfied or not at higher orders. Then we
impose the Petrov type \I  condition to reduce the gravity to the
dual non-relativistic hydrodynamics. With the ingoing Rindler metric
\begin{align}
\d s^2_{p+2} &= -r \d\t^2+2\d\t\d r+\d x_i\d x^i, \label{metricrindler}
\end{align}
the induced metric at the finite cutoff surface $r=r_c$ is
\begin{align}
\d s^2_{p+1} &=\gamma_{ab}\d x_a\d x^b= -r_c \d\t^2+\d x_i\d x^i.\label{imetriccut}
\end{align}
The Hamiltonian constraint becomes $\Ha=0$, where
\begin{align}\label{Hcut}
\Ha\equiv T^{\tau}_{~\tau}T^{\tau}_{~\tau}-2 r_c
\,T^{\tau}_{~i}\,T^{\tau}_{~j}\delta^{ij}+T_{\,ij}T^{\,ij}-p^{-1}T^2.
\end{align}
Defining $\Pe_{ij}=4C_{(\ell)i(\ell)j}$ and using equations (\ref{typeI}) and (\ref{BYcut}), the Petrov type
\I  condition turns out to be $\Pe_{ij}=0$, where
\begin{align}\label{petrovcut}
2 \Pe_{ij}\equiv&~{T^\tau}_\tau {T}_{\,ij} +
    {2 r_c}{T^\tau}_i{T^\tau}_j -
    {4 r_c^{-1/2}} \partial_{\tau}{T}_{\,ij} - {T}_{\,ik}{T^k}_{j} -
    {4 r_c^{1/2}}\partial_{(i}T^\tau_{~j)} \nonumber\\
    &+\,{p^{-2}}\left[T \left({T}-p {T^\tau}_\tau\right)+ 4 p r_c^{-1/2}\partial_\tau{T}\right]{\delta}_{\,ij}.
\end{align}

\subsection{Non-relativistic fluid and Petrov type \I  condition}
%{From non-relativistic fluid to the Petrov type \I  condition}
\label{Sec31}

Take the non-relativistic expansion in \cite{Bredberg:2011jq,Compere:2011dx}
\beq\label{orders}
v_i\sim \e,~~~~P \sim \e^2,~~~~\p_i \sim \e,~~~~\p_\tau \sim \e^2,
\eeq
the Brown-York stress tensor up to order $\eps^4$ can be expressed as \cite{Compere:2011dx}
\begin{align}
{T^\tau}_i=&+r_c^{-3/2}v_i+r_c^{-5/2}\[v_i(v^2+P)-2 r_c\s_{ij}v^j\]+O(\eps^5),\label{Tti}\\[1ex]
{T^\tau}_\tau=& -r_c^{-3/2}v^2 -r_c^{-5/2}\left[v^2(v^2+P)-2r_c \s_{ij}v^iv^j-{2
r_c^2}\s_{ij}\s^{ij}\right]+O(\eps^6),\label{Ttt} \\[1ex]
{T}_{\,ij}=&+ r_c^{-1/2}\,\delta_{ij}+ r_c^{-3/2}\[P\delta_{ij}+v_iv_j-2  r_c \s_{ij}\]\nn\\[1ex]
&+ r_c^{-5/2}\[ v_iv_j (v^2+P)-{  r_c}\s_{ij}v^2+2 r_c v_{(i}\p_{j)}P -  r_c v_{(i}\p_{j)}v^2-2 r_c^2
v_{(i}\p^2 v_{j)} \right.\nn\\[1ex]
&\left.-{2r_c^2}\s_{ik}{\s^k}_{j}-4 r_c^2{\s}_{k(i}\omega^k_{~j)}-4 r_c^2 \omega_{ik}{\omega^k}_{j}-4 r_c^2
\p_i\p_j P+{3r_c^3}\p^2 \s_{ij}\]+O(\eps^6),\label{Tij}\\[1ex]
{T}\ =&\,{T^\tau}_\tau+{T^i}_{i}= p\,r_c^{-1/2}+p\,r_c^{-3/2}P+O(\eps^6),\label{Tt}
\end{align}
where the fluid shear $\s_{ij}$ and vorticity $\omega_{ij}$ are given by
\footnote{Here the notations are different from \cite{Compere:2011dx} with a factor 2}
\be\label{sigmadef} \sigma_{ij}\equiv \D_{(i}v_{j)}= \(\D_iv_j+\D_jv_i\)/2, \qquad \omega_{ij}\equiv
\D_{[i}v_{j]} = \( \D_iv_j-\D_jv_i\)/2.
\ee
Comparing this stress tensor with the non-relativistic fluid stress tensor given in Appendix~\ref{B1}, one can
read off some transport coefficients as
\be\eta=1,\quad c_1= -2, \quad c_2=c_3=c_4=-4.
\ee
The equations of motion of the dual fluid $\p^aT_{ab}=0$ turn out to be the incompressible Navier-Stokes
equations with higher order corrections given in (\ref{ins}), and the stress tensor satisfies the Hamiltonian
constraint $\Ha=0$ consistently.
Inserting the stress tensor (\ref{Tti})-(\ref{Tt}) into $\Pe_{ij}$ and expanding in powers of parameter
$\epsilon$, one has
\begin{eqnarray}
\Pe_{ij}=\Pe_{ij}^{(0)}+\Pe_{ij}^{(2)}+\Pe_{ij}^{(4)}+O(\eps^6).
\end{eqnarray}
Taking into account the equations of motion (\ref{ins}),
one can see that $\Pe_{ij}^{(0)}$ and $\Pe_{ij}^{(2)}$ vanish identically, but
\begin{align}\label{Pij4}
\Pe_{ij}^{(4)}=&\,r_c^{-3}\[-6r_cv^kv_{(i}\omega_{j)k}-2r_c^{2}v_{(i}\partial^2v_{j)}+4
r_c^{2}v^k\partial_{(i}\omega_{j)k}+r_c^3\partial^2 \sigma_{ij}\].
\end{align}
This result can also be obtained through substituting the nonlinear solution of vacuum Einstein gravity given
in Appendix~\ref{A1} into the Weyl tensor (\ref{petrovij}) directly. And it is independent of the gauge
transformation that $v_i\rightarrow v_i+\delta v_i$ or $T_{ij} \rightarrow T_{ij} +\delta P\delta_{ij}$, where
$\delta v_i\sim\e^3,~ \delta P\sim\e^4$.
Thus the perturbed stress tensor  (\ref{Tti})-(\ref{Tt}) on the finite cutoff surface does not satisfy the
Petrov type \I  condition at order $\e^4$, if we choose this frame (\ref{frame}) associated with the finite
cutoff hypersurface. Or in other words, the non-linear solution of vacuum Einstein gravity constructed by boost
transformation, up to order $\epsilon^4$, does not satisfy the Petrov type \I  condition.

But we can additionally require the constraint $\Pe_{ij}^{(4)}=0$ holds. For example, if we take the
irroational condition with $\omega_{ij}\sim O(\eps^4)$, then in view of $\theta\equiv \p_i v^i \sim O(\eps^4)$,
one has
\begin{align}
\partial^2v_{j}=&\,\partial_j\theta-2\partial^k\omega_{jk}\sim O(\eps^5),\label{p2v} \\
\partial^2 \sigma_{ij}=&\,\partial_{(i}\partial_{j)}\theta-2\partial^k\partial_{(i}\omega_{j)k}\sim O(\eps^6).
\end{align}
Thus $\Pe_{ij}^{(4)}$ vanishes at this order and ${T}_{\,ij}$ is reduced to
\begin{align}
{T}^{(\sigma)}_{\,ij}=&\,\, r_c^{-1/2}\,\delta_{ij}+ r_c^{-3/2}\[P\delta_{ij}+v_iv_j-2  r_c \s_{ij}\]
+r_c^{-5/2}\[ v_iv_j (v^2+P)\right. \nn\\
&\left.-{r_c}\s_{ij}v^2 +2r_c v_{(i}\p_{j)}P -r_c v_{(i}\p_{j)}v^2
-{2r_c^2}\s_{ik}{\s^k}_{j}-4 r_c^2 \p_i\p_j P \].\label{Tijnon}
\end{align}
In this case, comparing (\ref{Tijnon}) with the non-relativistic fluid stress tensor in Appendix \ref{B1}, we
can read off
\be \eta=1,\quad c_1= -2, \quad c_4=-4.
\ee
The incompressible Navier-Stokes equations with higher order corrections (\ref{ins}) are reduced to
\be \label{ins3} \p_i v^i =\theta^{(\sigma)},
\qquad\p_\tau v_i +v^j \p_j v_i  +\p_i P = r_c \p^2 v_i+f^{(\sigma)}_i,
\ee
where the higher order corrections become
\begin{align}
\theta^{(\sigma)}=&  +{2}\s_{ij}\s^{ij}+{r^{-1}_c} v^i \p_i P + O(\ep^6),\\
f^{(\sigma)}_i=&-{3r_c}\p_i (\s_{kl}\s^{kl})+4 r_c \s^{kl} \p_k \s_{li}-2v^k \p_k \p_i P-2 (\p^k v_i)\p_k P \nn
\\
&- (\p_k \s_{il})v^k v^l+ r_c^{-1}(P+v^2) \p_i P -r_c^{-1}v_i\p_\tau P   + O(\eps^7)\, .\label{f3}
\end{align}
Here according to (\ref{p2v}), the term $r_c \p^2 v_i\sim O(\e^5)$, therefore we move this term to the right
hand side of the Navier-Stokes equations in (\ref{ins3}).

\subsection{From Petrov type \I  condition to dual fluid}
\label{Sec32}

At the finite cutoff surface, if we impose the Petrov type \I  condition $\Pe_{ij}=0$ firstly,
and consider the non-relativistic hydrodynamic scaling laws in (\ref{orders}), then the Brown-York stress
tensor can be expanded in powers of the non-relativistic hydrodynamic expansion parameter $\epsilon$ as
\begin{align}
    {T^\tau}_i     = &~{T^\tau}^{(1)}_i+ {T^\tau}^{(3)}_i + O(\eps^5),\nonumber\\
    {T^\tau}_\tau  = &~{T^\tau}^{(0)}_\tau + {T^\tau}^{(2)}_\tau+ {T^\tau}^{(4)}_\tau+ O(\eps^6),\nonumber\\
   {T}_{\,ij}     =&~{T}^{(0)}_{\,ij}+{T}^{(2)}_{\,ij}+{T}^{(4)}_{\,ij}+O(\eps^6),\nonumber\\
    T\,\,          = &~T^{(0)}+  T^{(2)} + T^{(4)} +O(\eps^6).\label{Tabepsilon}
\end{align}
Here superscript in round brackets stands for the expansion order, such as ${T^\tau}^{(1)}_i\sim\e,\
{T^\tau}^{(3)}_i\sim\e^3$, and so on. The Brown-York stress tensor at the cutoff surface $r=r_c$ of the metric
(\ref{metricrindler}) gives
\begin{equation}
~~~~~{T^\tau}^{(0)}_\tau=0,~~~~
{T}^{(0)}_{\,ij}=r_c^{-1/2}\,\delta_{ij},~~~
T^{(0)}=r_c^{-1/2}\,p\,.
\end{equation}
We now put the expansions (\ref{Tabepsilon}) into the Hamiltonian constraint equation (\ref{Hcut}) and the
Petrov equations (\ref{petrovcut}), which both can be expanded in powers of the parameter $\epsilon$. The first
non-trivial order appears at order $\e^2$, where the Hamiltonian constraint $\Ha^{(2)}=0$ and Petrov type \I
condition $\Pe^{(2)}_{ij}=0$ lead to
\begin{align}
{T^\tau}^{(2)}_\tau=&-{T^\tau}^{(1)}_i{T^\tau}^{(1)}_j\delta^{ij},\label{Ttt2H}\\
T_{\,ij}^{(2)} =&~p^{-1}{T^{(2)}}{\delta}_{ij}+r_c^{3/2}\, {{T^\tau}^{(1)}_i}{{T^\tau}^{(1)} _j}-2\,r_c\,
   \p_{(i}{T_{~j)}^{\tau(1)}}, \label{TijP2}
\end{align}
respectively. Following \cite{Lysov:2011xx}, if we assume that
\begin{eqnarray}
\quad{T^\tau}^{(1)}_i=r_c^{-3/2}v_i,\quad T^{(2)}=r_c^{-3/2}p\,P,\label{TtiP1}
\end{eqnarray}
we can recover the stress tensor (\ref{Tti})-(\ref{Tt}) up to order $\ep^2$.
The next non-trivial  Hamiltonian constraint $\Ha^{(4)}=0$ and Petrov type \I  condition $\Pe^{(4)}_{ij}=0$
give
\begin{align}
{T^\tau}^{(4)}_\tau=&-r_c^{3/2}{T^\tau}^{(1)}_i{T^\tau}^{(3)}_j\delta^{ij}+\frac{1}{2}
\[r_c^{1/2}{T}^{(2)}_{\,ij}{T}_{(2)}^{\,ij}
+ r_c^{1/2}({T^\tau}^{(2)}_\tau)^2 -{p^{-1}r_c^{1/2}}(T^{(2)})^2\],\label{Ttt4H}\\
T_{\,ij}^{(4)} =&~ 2\,r_c^{3/2} {{T^\tau}^{(1)}_{(i}}{{T^\tau}^{(3)}_{j)}} -2
    r_c\,\p_{(i}T_{~j)}^{\tau(3)}+{1\over{2}}r^{1/2}_c{{T^\tau}^{(2)}_\tau}T^{(2)}_{\,ij}-{1\over{2}}T^{(2)}_{\,ik}{T^{(2)}_{\,j\,l}}\delta^{kl}
-\p_\tau {T^{(2)}_{\,ij}}\nn\\
&+\frac{1}{2}\,p^{-1}\[p^{-1}r_c^{1/2}(T^{(2)})^2-r_c^{1/2}{T^{(2)}{{T^\tau}^{(2)}_\tau}}+{4\p_\tau
T^{(2)}}+{2\,T^{(4)}}\]{\delta}_{\,ij},\label{TijP4}
\end{align}
respectively. To give assumptions at higher orders, we choose the Landau frame which gives
\be 0=h_a^{\,b}T_{bc}u^c,\quad h_a^{b}=\delta_a^{\,b}+u_a u^b, \ee
where $u^a=\gamma_v(1,v^i)$ and $\gamma_{ab}u^au^b=-1$~\cite{Compere:2011dx}. At order $\e^3$, its spatial
components lead to
\be 0=-r_cT^{\tau(3)}_{~i}+T^{(2)}_{\,ij}v^j+\erho^{(2)}v_i, \ee
where the energy density $\erho\equiv T_{ab}u^au^b$. With the recovered stress tensor up to $\e^2$, one can
show $\erho^{(2)}=0$. Putting (\ref{TijP2}) and (\ref{TtiP1}) into the above equation, we obtain
\begin{eqnarray}\label{TtiP3} {T^\tau}^{(3)}_i=r_c^{-5/2}\[v_i(v^2+P)-2 r_c\s_{ij}v^j\].
\end{eqnarray}
Then ${T^\tau}_\tau$ in (\ref{Ttt}) can be recovered up to order $\e^4$ with the Hamiltonian constraint which
leads to (\ref{Ttt2H}) and (\ref{Ttt4H}).
On the other hand,  putting (\ref{TijP2}) (\ref{TtiP1}) and (\ref{TtiP3}) into (\ref{TijP4}),
one finds that %${T}^{(2)}_{\,ij}$ is the same as terms in (\ref{Tij}) at order $\e^2$, and
at order $\e^4$, there is only one term $T^{(4)}\delta_{ij}$ proportional to $\delta_{ij}$.
Thus, we can choose the isotropic gauge with $T^{(4)}=0$ as in~\cite{Compere:2011dx}, and finally
${T}^{(4)}_{\,ij}$ is given by
\begin{align}
{T}^{(4)}_{\,ij}=&\, r_c^{-5/2}\[ v_iv_j (v^2+P) -{r_c}\s_{ij}v^2+2r_c v_{(i}\p_{j)}P -r_c v_{(i}\p_{j)}v^2
+6r_cv_kv_{(i}\omega^k_{~j)} -4r_c^2 v_{(i}\p^2 v_{j)} \right.\nn\\[1ex]
&\left.-{2r_c^2}\s_{ik}{\s^k}_{j}-4 r_c^2{\s}_{k(i}{\omega^{k}_{~j)}}-4 r_c^2 \omega_{ik}{\omega^k}_{j}-4 r_c^2
\p_i\p_j P-4 r_c^{2}v_k\partial_{(i}\omega^k_{~j)}+{4r_c^3}\p^2 \s_{ij}\].\label{T4ijP4}
\end{align}
Compare (\ref{T4ijP4}) with the terms in (\ref{Tij}) at order $\e^4$, we obtain the additional terms
\begin{align}\label{Pij4T}
\,r_c^{-5/2}\[6r_cv_kv_{(i}\omega^k_{~j)}-2r_c^{2}v_{(i}\partial^2v_{j)}-4
r_c^{2}v_k\partial_{(i}\omega^k_{~j)}+r_c^3\partial^2 \sigma_{ij}\].
\end{align}
Thus, the incompressible Navier-Stokes equations with higher order corrections from the equations of motion of
the fluid $\p^a T_{ab}=0$ become
\be \label{inso} \p_i v^i =\theta,
\qquad\p_\tau v_i +v^j \p_j v_i -r_c \p^2 v_i +\p_i P = f_i+f^{(\omega)}_i,
\ee
where $\theta$ and $f_i$ are given in (\ref{theta}) and (\ref{fi}), respectively, and
\begin{align}
f^{(\omega)}_i=&-\frac{r_c^2}{2}\p^4 v_i+4r_cv^k \p^2 \omega_{ki}+2r_c  \p_l \omega_{ki} \p^l
v^k+{2r_c}\p_kv_i\p_l \omega^{lk}+r_c\p_i (\omega_{kl}\omega^{lk}) \nn \\
&- 3v_i(\omega_{kl}\omega^{lk})-3v_iv_k\p_l\omega^{kl}-3v_k\omega_{li}\p^k v^l -3v_k(\p_l
v_i)\omega^{kl}-3(\p_l \omega_{ki})v^k v^l   + O(\eps^7) .\label{fomega}
\end{align}

Comparing (\ref{T4ijP4}) with the non-relativistic fluid dual to vacuum Einstein gravity constructed in
Appendix \ref{B1}, one can extract the second order transport coefficients as
\be \label{2ndcoefficients} c_1= -2, \quad c_2=c_3=c_4=-4,
\ee
which implies that the correction terms in (\ref{Pij4}) do not contribute to the terms associated with second
order transport coefficients. Thus, such kind of higher order fluid reduced from the Petrov type \I  condition,
which we name as ``Petrov type \I  fluid'',
does not satisfy the non-relativistic fluid that constructed in Appendix \ref{B1}.
However, if additionally requiring that the terms in (\ref{Pij4}) vanish at this order, we can again recover
the previous stress tensor (\ref{Tti})-(\ref{Tt}), up to order $\epsilon^4$. In particular, taking the
irrotational condition that $\omega_{ij}\sim O(\e^4)$, we can recover equations (\ref{Tijnon})-(\ref{f3}).

\section{On highly accelerated surface}
%{Near horizon expansion}
\label{Sec4}

An alternative presentation of the procedure discussed in the previous section can also be realized with the
near horizon expansion.
Introducing the expansion parameter $\lambda=r_c^{1/2}$ via the transformation $\tau \rightarrow\lambda^{-2}
\htau,~r\rightarrow\l^{2} \hat r,~x\rightarrow\x$, the ingoing Rindler metric (\ref{metricrindler}) becomes
\begin{align}\label{metricnear}
\d \hat s^2_{p+2} &= -\frac{\r}{\l^2} \d\htau^2+2\d\htau\d \r+\d \x_i\d \x^i,
\end{align}
which gives the first three terms in (\ref{metricnear1}). The induced metric (\ref{imetriccut}) changes into
\begin{align}\label{imetricnear}
\d \hat s^2_{p+1} &=\hat\gamma_{ab}\d\x^a\d\x^b= -\frac{1}{\l^2} \d\htau^2+\d \x_i\d \x^i.
\end{align}
In the hatted coordinates, the Hamiltonian constraint becomes $\hHa=0$, where
\begin{align}\label{hnear}
\hHa\equiv\hT^{\htau}_{~\htau}\hT^{\htau}_{~\htau}-2\l^{-2}\hT^{\htau}_{~i}\hT^{\htau}_{~j}\delta^{ij}+\hT_{\,ij}\hT^{\,ij}-p^{-1}\hT^2.
\end{align}
The Petrov type \I  condition turns out to be $\hPe_{ij}=0$, where
\begin{align}\label{petrovnear}
2\hPe_{ij}\equiv&~{\hT^\htau}_{~\htau} {\hT}_{\,ij} +
    {2\l^{-2}}{\hT^\htau}_{~i}{\hT^\htau}_{~j} -
    {4 \l\,} \hpartial_{\htau}{\hT}_{\,ij} - {\hT}_{\,ik}{\hT^k}_{~j} -
    {4 \l^{-1}}\hpartial_{(i}{T^\htau}_{j)} \nonumber\\
    &+\,{p^{-2}}\left[\hT({\hT}-p {\hT^\htau}_{~\htau})+4p \l\,\partial_\htau{\hT}\right]{\delta}_{\,ij}.
\end{align}

\subsection{Near horizon fluid and Petrov type \I  condition}
%{From the near horizon fluid to the Petrov type \I  condition}
\label{Sec41}

In the near horizon expansion, with the transformations (\ref{hatx}),(\ref{hatP}) and (\ref{nearBY}), the
stress tensor (\ref{Tti})-(\ref{Tt}) becomes
\begin{align}
{\hT^\htau}_{~i}=&+\lambda v_i+\lambda^3\[\v_i(\v^2+\P)-2 \hs_{ij}\v^j\]+O(\l^5),\label{hTti}\\[1ex]
{\hT^\htau}_{~\htau}=& -\l v^2
-\l^3\left[\v^2(\v^2+\P)-2\hs_{ij}\v^i\v^j-2\hs_{ij}\hs^{ij}\right]+O(\l^5),\label{hTtt} \\[1ex]
{\hT}_{\,ij}=&+ \l^{-1}\,\delta_{ij}+ \l\[\P\delta_{ij}+\v_i\v_j-2  \hs_{ij}\]\nn\\[1ex]
&+ \l^{3}\[ \v_i\v_j (\v^2+\P)-\hs_{ij}\v^2+2 \v_{(i}\hp_{j)}\P - \v_{(i}\hp_{j)}\v^2-2\v_{(i}\hp^2 \v_{j)}
\right.\nn\\[1ex]
&\left.-{2}\hs_{ik}{\hs^k}_{~j}-4 {\hs}_{k(i}\homega^{k}_{~j)}-4 \homega_{ik}{\homega^k}_{~j}-4 \hp_i\hp_j
\P+{3}\hp^2 \hs_{ij}\]+O(\l^5),\label{hTij}\\[1ex]
{\hT}\ =&\,{\hT^\htau}_{~\htau}+{\hT^i}_{~i}= \l^{-1}p+\l\,p\,P+O(\l^5),\label{hTt}
\end{align}
where the fluid shear $ \hsigma_{ij}\equiv \hp_{(i}\v_{j)}$ and vorticity $\homega_{ij}\equiv \hp_{[i}\v_{j]}$.
Comparing the stress tensor with the one of dual fluid given in Appendix \ref{B2}, one has
\be \hat\eta=1,\quad \hc_1= -2, \quad \hc_2=\hc_3=\hc_4=-4.
\ee
The equations of motion $\hp^a\hT_{ab}=0$ turn out to be (\ref{hins}), and the stress tensor satisfies the
Hamiltonian constraint $\hHa=0$  consistently.
Inserting equations (\ref{hTti})-(\ref{hTt}) into $\hPe_{ij}$ with expansion in powers of $\l$, we have
\begin{eqnarray}
\hPe_{ij}=\l^{-2}\hPe_{ij}^{(-2)}+\l^{0}\hPe_{ij}^{(0)}+\l^{2}\hPe_{ij}^{(2)}+O(\l^4).
\end{eqnarray}
We see that $\hPe_{ij}^{(-2)}$ and $\hPe_{ij}^{(0)}$ vanish identically, but
\begin{align}
\hPe_{ij}^{(2)}=&\,-6\v^k\v_{(i}\homega_{j)k}-2\v_{(i}\hpartial^2\v_{j)}+4 \v^k\hpartial_{(i}\homega_{j)k}+
\hpartial^2 \hsigma_{ij}.
\end{align}
This is independent of the gauge transformation with $\v_i\rightarrow \v_i+\l^2\delta \v_i$ or $\hT_{ij}
\rightarrow \hT_{ij} +\l^3\delta \P\delta_{ij}$.
Thus the perturbed stress tensor  (\ref{hTti})-(\ref{hTt}) does not satisfy the Petrov type \I  condition at
order $\l^2$, if we choose this frame (\ref{frame}).

Again, we can also additionally require  $\hPe_{ij}^{(2)}=0$. For example, if we add the irroational condition
that $\homega_{ij}\sim O(\l^2)$, then $\hPe_{ij}^{(2)}$ vanishes at this order and ${\hT}_{\,ij}$ is reduced to
\begin{align}\label{hTijnon}
{\hT}^{(\hsigma)}_{\,ij}=&~\l^{-1}\,\delta_{ij}+ \l\[\P\delta_{ij}+\v_i\v_j-2 \hs_{ij}\]
+\l^{3}\[ \v_i\v_j (\v^2+\P)\right. \nn\\
&\left.-\hs_{ij}\v^2 +2\v_{(i}\hp_{j)}\P -\v_{(i}\hp_{j)}\v^2
-{2}\hs_{ik}{\hs^k}_{j}-4 \hp_i\hp_j P \].
\end{align}
Comparing this with the stress tensor of dual fluid given in Appendix  \ref{B2}, we have
\be \heta=1,\quad \hc_1= -2, \quad \hc_4=-4.
\ee
In this case, the incompressible Navier-Stokes equations with higher order corrections (\ref{hins}) are reduced
to
\be \label{hins3}\hp_i \v^i =\htheta^{(\hsigma)}, \quad\p_\htau \v_i +\v^j \hp_j \v_i  +\hp_i \P = \hp^2
\v_i+\hf^{(\hsigma)}_i,
\ee
where the higher order corrections are given by
\begin{align}
\htheta^{(\hsigma)}=&\, \l^2 \[+{2}\hs_{ij}\hs^{ij}+ \v^i \hp_i \P\] + O(\l^4),\\
\hf^{(\hsigma)}_i=&\,\l^2 \[-{3}\hp_i (\hs_{kl}\hs^{kl})+4 \hs^{kl} \hp_k \hs_{li}-2\v^k \hp_k \hp_i \P-2
(\hp^k \v_i)\hp_k \P\right. \nn \\
&\left.\quad~- (\hp_k \hs_{il})\v^k \v^l+ (\P+\v^2) \hp_i \P -\v_i\hp_\htau \P \]  + O(\l^4)\, .\label{hf3}
\end{align}
Since the term $\hp^2 \v_i\sim O(\l^2)$, it is therefore put on the right hand side of the equation
(\ref{hins3}).

\subsection{From Petrov type \I  condition to dual fluid}
\label{Sec42}

In this subsection we will inverse the procedure and expand the Brown-York stress tensor in powers of the
parameter $\lambda$ with the background metric (\ref{imetricnear}),
\begin{align}\label{hTabepsilon}
    {{\hT}^\htau}_{~i}& =  \lambda\,{\hT}^{\htau(1)}_{~i} + \lambda^3\,{\hT}^{\htau(3)}_{~i}+ O(\l^5),\nn\\
    {\hT^\htau}_{~\htau}&  =  \lambda\,{\hT}^{\htau(1)}_{~\htau}+\lambda^3\,{\hT}^{\htau(3)}_{~\htau}+
    O(\l^5),\nn\\
    {\hT}_{\,ij}&        = \l^{-1}\delta_{\,ij} + \lambda\,{\hT^{(1)}_{\,ij}}+ \lambda^3\,{\hT^{(3)}_{\,ij}}+
    O(\l^5),\nn\\
    \hT \,\,&          = \l^{-1}p + \lambda\, \hT^{(1)} + \lambda^3\, {\hT}^{(3)} + O(\l^5).
\end{align}
Note that here only the odd order terms are selected. The even order
terms can also be added, because it can be shown that they give no
further information of the higher order fluid, and thus are set to
be vanished to satisfy the constraint equations as well as Petrov
type \I  condition. We now put the expansions (\ref{hTabepsilon})
into the Hamiltonian equation (\ref{hnear}) and the Petrov equations
(\ref{petrovnear}), which both can be expanded in powers of the
parameter $\l$. The first non-trivial order appears at $\l^0$, where
the Hamiltonian constraint $\hHa^{(0)}=0$ and Petrov type \I
condition $\hPe^{(0)}_{ij}=0$ lead to \begin{align}
\hT^{\htau(1)}_{~\htau}=&-\hT^{\tau(1)}_{~i}\hT^{\tau(1)}_{~j}\delta^{ij},\label{hTttH1}\\
\hT_{\,ij}^{(1)} =&~p^{-1}{\hT^{(1)}}{\delta}_{ij}+\, \hT^{\htau(1)}_{~i}\hT^{\htau(1)}
_{~j}-2\,\hp_{(i}{\hT_{~j)}^{\htau(1)} },\label{hTijP2}
\end{align}
respectively. Again, following \cite{Lysov:2011xx}, if assuming that
\begin{eqnarray}
\hT^{\tau(1)}_{~i}=\v_i,\qquad \hT^{(1)}=p\,\P,\label{hTtiP1}
\end{eqnarray}
we can recover the stress tensor (\ref{hTti})-(\ref{hTt}) up to order $\l$. The next non-trivial Hamiltonian
constraint $\hHa^{(2)}=0$ and Petrov type \I  condition $\hPe^{(2)}_{ij}=0$ give
\begin{align}
\hT^{\htau(3)}_{~\htau}=&-\hT^{\tau(1)}_{~i}\hT^{\tau(3)}_{~j}\delta^{ij}+\frac{1}{2}
\[{\hT}^{(1)}_{\,ij}{\hT}_{(1)}^{\,ij}
+ (\hT^{\htau(1)}_{~\htau})^2 -{p^{-1}}(\hT^{(1)})^2\],\label{hTttH3}\\
\hT_{\,ij}^{(3)} =&~2\,\hT^{\tau(1)}_{~(i}\hT^{\tau(3)}_{~j)}-2\,\hp_{(i}\hT_{~j)}^{\htau(3)}
+{1\over{2}}\hT^{\htau(1)}_{~\htau}\hT^{(1)}_{\,ij}-{1\over{2}}{{\hT}^{(1)}_{i\,k}\hT^{(1)}_{\,j\,l}}\delta^{kl}
-2\hp_\htau {\hT^{(1)}_{\,ij}}\nn\\
&+\frac{1}{2}\,p^{-1}\[p^{-1}(\hT^{(1)})^2-\hT^{(1)}\hT^{\htau(1)}_{~\tau}+4\hp_\htau
\hT^{(1)}+{2\,\hT^{(3)}}\]{\delta}_{\,ij},\label{hTijP4}
\end{align}
respectively. To give assumptions at higher order, we choose the Landau frame which gives
\be 0=\hat h_a^{\,b}\hat T_{bc}\hat u^c,\qquad \hat h_a^{b}=\delta_a^{\,b}+\hat u_a \hat u^b,
\ee
where $\hat u^a=\hat\gamma_v(1,\v^i)$ and $\hat\gamma_{ab}\hat u^a\hat u^b=-1$.
At order $\l$, the spatial components give us with
\be\label{hLandau} 0=-\hT^{\htau(3)}_{~ i}+\hT^{(1)}_{\,ij}\v^j+\hat\erho^{(1)}v_i,
\ee
where $\hrho\equiv\hT_{ab}\hat u^a\hat u^b$. From the recovered stress tensor up to order $\l$ we have
$\erho^{(1)}=0$.
Putting (\ref{hTijP2}) and (\ref{hTtiP1}) into the above equation we get
\begin{eqnarray}
\hT^{\tau(3)}_{~i}=\v_i(\v^2+\P)-2 \hs_{ij}\v^j.\label{hTtiP3}
\end{eqnarray}
Then $\hT^{\htau}_{~\tau}$ in (\ref{hTtt}) can be recovered up to
order $\l^3$ via the Hamiltonian constraint which leads to
(\ref{hTttH1}) and (\ref{hTttH3}). On the other hand, putting
(\ref{hTijP2})(\ref{hTtiP1}) and (\ref{hTtiP3}) into (\ref{hTijP4}),
one finds that at order $\l^3$, there is only one term
$\hT^{(3)}\delta_{ij}$ proportional to $\delta_{ij}$. Thus, we can
choose the isotropic gauge so that $\hT^{(3)}=0$ and
$\hT^{(3)}_{\,ij}$ can be expressed as
\begin{align}\label{hT4ijP4}
\hT^{(3)}_{\,ij}=&\,\v_i\v_j (\v^2+\P) -\hs_{ij}\v^2+2\v_{(i}\hp_{j)}\P -\v_{(i}\hp_{j)}\v^2
+6\v_k\v_{(i}\homega^k_{~j)} -4 \v_{(i}\hp^2 \v_{j)}\nn\\[1ex]
&-{2}\hs_{ik}\hs^k_{~j}-4 \hs_{k(i}\homega^{k}_{~j)}-4 \homega_{ik}{\homega^k}_{~j}-4 \hp_i\hp_j \P-4
\v_k\hpartial_{(i}\homega^k_{~j)}+{4}\hp^2 \hs_{ij}.
\end{align}
Comparing (\ref{hT4ijP4}) with the terms in (\ref{hTij}) at order $\l^3$, one can find that the additional
terms are
\begin{equation}\label{hPij4}
6\v_k\v_{(i}\homega^k_{~j)}-2\v_{(i}\hpartial^2\v_{j)}-4 \v_k\hpartial_{(i}\homega^k_{~j)}+ \hpartial^2
\hsigma_{ij}.
\end{equation}
Thus, the incompressible Navier-Stokes equations with higher order corrections from the equations of motion of
the fluid $\hp^{ a} \hT_{a b}=0$ become
\be \label{hoins} \hp_i \v^i =\htheta, \quad\p_\htau \v_i +\v^j \hp_j \v_i -\hp^2 \v_i +\hp_i \P =
\hf_i+\hf^{(\homega)}_i,
\ee
where $\htheta$ and $\hf_i$ are given in (\ref{htheta}) and (\ref{hfi}), respectively, and
\begin{align}
\hf^{(\homega)}_i=&\l^2\[-\frac{1}{2}\hp^4 \v_i+4\v^k \hp^2 \homega_{ki}+2\hp_l \homega_{ki} \hp^l
\v^k+{2}\hp_k\v_i\hp_l \homega^{lk}+\hp_i (\homega_{kl}\homega^{lk}) -
3\v_i(\homega_{kl}\homega^{lk})\right.\nn \\
&\left.\quad~-3\v_i\v_k\hp_l\homega^{kl}-3\v_k\homega_{li}\hp^k \v^l -3\v_k(\hp_l \v_i)\homega^{kl}-3(\hp_l
\homega_{ki})\v^k \v^l \]  + O(\l^4) .\label{hfomega}
\end{align}
Comparing (\ref{hT4ijP4}) with the stress tensor of fluid given in Appendix \ref{B2}, one can obtain the second
order transport coefficients as
\be \label{2ndhc} \hc_1= -2, \quad \hc_2=\hc_3=\hc_4=-4. \ee
Thus, we have shown that the additional corrections do not make contribution to the second order transport
coefficients. Such kind of higher order Petrov type \I  non-relativistic fluid does not match  the fluid
constructed in Appendix \ref{B2}. However, if we additionally require that the terms in (\ref{hPij4}) vanishes
at this order, the stress tensor (\ref{hTti})-(\ref{hTt}) can be recovered.
In particular, taking the irrotational condition with $\homega_{ij}\sim O(\l^2)$, we can still recover
equations (\ref{hTijnon})-(\ref{hf3}).

\section{Conclusion and Discussion}
\label{Sec5}

In Einstein gravity, the Petrov type \I  condition relates the
gravity theory to a dual fluid without gravity in one lower
dimension. It reduces the the extrinsic curvature of a time-like
hypersurface to $p+2$ components, which can be interpreted as the
energy density, pressure and velocity field of a dual
fluid living on the hypersurface, constrained by  equation of state
and $p+1$ evolution equations (incompressible Navier-Stokes
equations) that come from the Einstein constraint
equations~\cite{Lysov:2011xx}. To the leading order there are two
equivalent presentations, one is for the dual fluid living on a
finite cutoff surface via non-relativistic hydrodynamic expansion,
and the other on a highly accelerated surface via the near horizon
expansion. Imposing the Petrov condition is mathematically much
simpler than imposing regularity on the future horizon.

Via appropriate gauge choice, we generalized this procedure to the next order and obtained the
incompressible Navier-Stokes equations with higher order corrections
and associated second order transport coefficients.
More higher order hydrodynamics can also be obtained order by order with appropriate expansion parameters.
We can recover
the non-relativistic fluid stress tensor dual to vacuum Einstein gravity
from boost transformation up to order $\epsilon^4$, only if by imposing additional constraint
such as the irrotational condition. In other words,  the non-linear
solution of vacuum Einstein equations constructed by boost
transformation does not satisfy the Petrov type \I  condition up to order
$\epsilon^4$, although it holds at the order $\epsilon^2$.

As the dual fluid constructed in Appendix \ref{AB} is reduced from
the relativistic hydrodynamics, while the Petrov type \I  condition
singles out a preferred time coordinate and thus breaks Lorentz
invariance of the hypersurface~\cite{Lysov:2011xx}, it might
be not surprised that the ``Petrov type \I  fluid'' does not match
the boosted fluid at higher orders. In this sense it would be
interesting to construct the non-relativistic hydrodynamics of this
special higher order fluid directly, with the corresponding
non-linear gravitational solution. In addition, note that the Petrov type \I  condition (\ref{petrovij}) looks
different depending on choice of frame. In this sense it should be instructive to consider a different frame
instead of (\ref{frame}), such as $m_i=h_i,\sqrt{2}\ell =u-n,\sqrt{2}k =-u-n$, where $u$ is the fluid velocity
and $h_i$ (with $i=1,...,p$) are the spatial basses orthogonal to both $u$ and $n$.\,\footnote{We are grateful
to the referee's valuable suggestions on this point. In fact this frame has been used in an ongoing work for
the dual relativistic fluid~\cite{Cai:2014ywa}.} %
In general, these basses can be written as
\begin{align} \label{boostframe}
m_i=&\p_i+\g \beta_i\p_0+(\g-1)\beta^{-2}\beta_i \beta^j\p_j,\nn\\
\sqrt{2}\ell =&\g(\p_0+\beta^i \p_i)-n^\mu\p_\mu,\nn \\
\sqrt{2}k=&-\g(\p_0+\beta^i\p_i)-n^\mu\p_\mu,
\end{align}
where the fluid velocity $u$ has be defined as $(\g,\g\b_i)$ with $\g=(1-\beta^2)^{-1/2}$, and $\beta_i\equiv
r_c^{-1/2}v_i$ if we use the induced metric (\ref{imetriccut}) on a finite cutoff surface. Taking the
non-relativistic hydrodynamic limit in (\ref{orders})
one can show that the result in (\ref{Pij4}) becomes $\Pe_{ij}^{(4)}=\partial^2 \sigma_{ij}$, which is the
third order in derivative expansion of the velocity. Inversely, if the condition $\Pe_{ij}=0$ is imposed, only
one term $f_i^{(\omega)}=-\frac{r_c^2}{2}\p^4 v_i$ is left in the correction terms (\ref{fomega}).

Thus, the frame given in (\ref{boostframe}) has better properties than the one in (\ref{frame}),
and it is expected that in the case of derivative expansion with the relativistic fluid
solution~\cite{Compere:2012mt,Eling:2012ni}, the Petrov type \I  condition in this frame holds at least up to
second order~\cite{Cai:2014ywa}.
In the non-relativistic hydrodynamic expansion discussed in this paper, the additional term
$\Pe_{ij}^{(4)}=\partial^2 \sigma_{ij}$ %that at order $\e^4$
might be reduced from the third order in derivative expansion of the relativistic fluid. In the near horizon
expansion, the situation is similar.  Thus it would be an interesting question whether one can find a frame
such that the Petrov type \I  condition on finite cutoff surface holds up to order $\e^4$.
In addition,  changing the boundary conditions of the
hypersurface to see their effects at higher orders, and generalizing
to other bulk geometries would also be valued for further works.

{\bf Note added:} During the preparation of this work, we were
informed that the leading order calculation in section \ref{Sec32}
might have some overlap with the work %in progress by authors
in~\cite{Ling:2013kua}. After finishing this work, we were told that
the authors in~\cite{Compere:2011dx} also reached  the conclusion
that the Petrov type \I  condition does not hold at higher orders
for the non-linear solution of vacuum Einstein gravity (unpublished,
May 2011).

\section*{Acknowledgements}
This work was supported in part by the National Natural Science
Foundation of China (No.10821504, No.10975168 and No.11035008), and
in part by the Ministry of Science and Technology of China under
Grant No. 2010CB833004. We thank Professors Yi Ling, Yu Tian and
Xiao-Ning Wu for useful discussions. In particular, we thank
Professor Kostas Skenderis for helpful correspondence and valuable
comments on this manucript. R. -G. Cai thanks the long-term workshop
YITP-T-12-03 on ``Gravity and Cosmology 2012'', Y. -L. Zhang  thanks
the  workshop on ``Gauge/Gravity Duality'' (YIPQS (2012)). Both
workshops were held at Yukawa Institute for Theoretical Physics at
Kyoto University.

\appendix

\section{Nonlinear metric solution}
\label{AA}
In this Appendix, we briefly give the nonlinear solution of vacuum Einstein equations $G_{\mu\nu}=0$, which is
obtained via the non-relativistic hydrodynamic expansion and near horizon expansion,
respectively~\cite{Bredberg:2011jq,Compere:2011dx}.

\subsection{Non-relativistic hydrodynamic expansion}
\label{A1}
Associated with the non-relativistic hydrodynamic expansion
\beq\label{ordersA1}
v_i\sim \e,\quad P \sim \e^2,\quad \p_i \sim \e,\quad \p_\tau \sim \e^2,\quad \p_r \sim \e^0,
\eeq
the metric which solves Einstein equations (\ref{Einstein}) up to order $\e^4$ is given as
\cite{Compere:2011dx},
\begin{align}
\label{aseed}
\d s^2_{p+2} &= -r \d\t^2+2\d\t\d r+\d x_i\d x^i
-2\big(1-\frac{r}{r_c}\big)v_i\d x^i \d \t-\frac{2}{r_c}v_i\d x^i \d r \nn\\
&\quad +\big(1-\frac{r}{r_c}\big)\Big[(v^2+2P)\d\t^2+\frac{1}{r_c}v_iv_j\d x^i\d x^j\Big]+\frac{1}{r_c}\(v^2+
2P\)\d\t\d r\nn\\
&\quad +2g^{(3)}_{\t i}\d x^i \d \t+g^{(4)}_{\tau\tau}\d \tau^2+g^{(4)}_{\,ij}\d x^i \d x^j+O(\ep^5),
\end{align}
where
\begin{align}
 g^{(3)}_{\t i} =& \,\frac{(r-r_c)}{2r_c}\Big[(v^2+2P)\frac{2v_i}{r_c}+4\D_iP -(r+r_c)\D^2 v_i\Big],\nn\\
g^{(4)}_{\tau\tau} =& -\frac{(r-r_c)^3}{2r_c^2}(\omega_{kl}\omega^{kl}) +\frac{(r-r_c)^2}{2 r_c}( 2 v^k\p^2 v_k
+ \s_{kl}\s^{kl})-\frac{(r-r_c)}{r_c} F_{\tau}^{(4)},\nn\\
F^{(4)}_{\tau} =& \,\frac{9}{8r_c}v^4+\frac{5}{2r_c}Pv^2+\frac{P^2}{r_c}-2r_c v^k \p^2 v_k -
2r_c\s_{kl}\s^{kl}-2\partial_\tau P + 2v^k \partial_k P,\nn\\
g^{(4)}_{ij} = &\,(1-\frac{r}{r_c}) \Big[ \frac{1}{r_c^2}\(v_i v_j-r_c\s_{ij}\) (v^2+2P)
 +\frac{2}{r_c}v_{(i}\p_{j)}P -\frac{1}{r_c}v_{(i}\p_{j)}v^2-\frac{r+r_c}{r_c}v_{(i}\p^2 v_{j)}\nonumber\\
&\qquad\qquad -{2}\s_{ik}\s^k_{~j}- 4\s_{k(i}\omega_{~j)}^{k}+\frac{r-5r_c}{r_c}\omega_{ik}\omega^k_{~j}
-4\p_i\p_j P +\frac{r+5r_c}{2}\p^2\s_{ij}\Big]. \label{gij4}
\end{align}
The dual fluid satisfies the incompressible Navier-Stokes equations with higher order corrections
\be
\label{ins}
\p_i v^i =\theta,
\qquad\p_\tau v_i +v^j \p_j v_i -r_c \p^2 v_i +\p_i P = f_i,
\ee
where
\begin{align}
\theta=&  -v^i \p^2 v_i +{2}\s_{ij}\s^{ij}+\frac{1}{r_c} v^i \p_i P + O(\ep^6),\label{theta}\\
f_i=&-\frac{3r_c^2}{2}\p^4 v_i+2r_cv^k \p^2 \p_k v_i+4r_c \s_{ik}\p_l \s^{kl}-{10r_c}\omega_{ik}\p_l
\s^{kl}-{3r_c}\p_i (\s_{kl}\s^{kl}) \nn \\
&- \frac{5r_c}{2}\p_i (\omega_{kl}\omega^{lk})+4 r_c \s^{kl} \p_k \s_{li}-2v^k \p_k \p_i P-2 (\p^k v_i)\p_k
P-(P+\frac{1}{2}v^2) \p^2 v_i  \nn \\
&- (\p_k \s_{il})v^k v^l+(\p_k \omega_{il}) v^k v^l+4(\p_k v_i )\omega^{kl}v_l +r_c^{-1}(P+v^2) \p_i P -
{r_c^{-1}}v_i\p_\tau P   + O(\eps^7) .\label{fi}
\end{align}

\subsection{Near horizon expansion}
\label{A2}
An alternate presentation of the  metric (\ref{aseed}) was given in \cite{Bredberg:2011jq}, through taking the
coordinate transformations
\beq\label{hatx}
{\x^i }  =\e\,{r_c^{-1}x^i},\quad \htau= \e^2\,r^{-1}_c\tau,\quad \r= r^{-1}_c r,
\eeq
so that %\beq\label{hatp}
$\hat \p_i\equiv{\p \over \p \x^i}\sim\e^0,~\p_{\htau }\sim\e^0,$ and $\p_{\r}\sim\e^0.$ %\eeq
In the new coordinates one defines
\beq\label{hatP}
\P(\x,\htau)=\e^{-2}P(x(\x),\tau(\htau)), \quad \v_i(\x,\htau)= \e^{-1}v_i(x(\x),\tau(\htau)),
\eeq
and $\v^2\equiv \v_i\delta^{ij}\v_j$.
Considering the rescaled metric ${\d\hat s}_{p+2}^2={\e^2 \,r_c^{-2} } \d s_{p+2}^2$ and defining $ \lambda^2={
\e^2}\, {r^{-1}_c}$, one finds
\begin{align}
\d\hat s_{p+2}^2 =&-{\r \over \lambda^2} \d\htau^2 %\notag\\&
+\bigl[2\d\htau \d\r + \d\x_i \d\x^i-2 (1-\r)\v_i \d\x^i \d\htau +{(1-\r)} ( \v^2+2\P) \d\htau^2 \bigr]  \nn\\
&+\lambda^2\bigl[(1-\r){\v_i \v_j }\d\x^i \d\x^j   -2 {\v_i}\d\x^i \d\r + (\v^2+2\P)\d\htau
\d\r+2{\hat{g}}^{(2)}_{\htau i} \d\x^i \d\htau +  {\hat{g}}^{(2)}_{\htau\htau} \d\htau^2 \bigr]\nn\\
&+\l^4\,\[{\hat{g}}^{(4)}_{ij}\d\x^i \d\x^j+2{\hat{g}}^{(4)}_{\htau i} \d\x^i \d\htau +
{\hat{g}}^{(4)}_{\htau\htau} \d\htau^2\]+O(\l^6),\label{metricnear1}
\end{align}
where
\begin{align}
 {\hat{g}}^{(2)}_{\htau i} =& \,\frac{(\r-1)}{2}\Big[(\v^2+2\P){2\v_i}+4\D_i\P -(\r+1)\hp^2 v_i\Big],\nn\\
\hat{g}^{(2)}_{\htau\htau} =& -\frac{(\r-1)^3}{2}(\homega_{kl}\homega^{kl}) +\frac{(\r-1)^2}{2}( 2 \v^k\hp^2
\v_k + \hs_{kl}\hs^{kl})-{(\r-1)} \hF_{\htau}^{(2)},\nn\\
\hF^{(2)}_{\htau} =& \,\frac{9}{8}\v^4+\frac{5}{2}\P\v^2+{\P^2}-2\v^k \hp^2 \v_k -
2\hs_{kl}\hs^{kl}-2\hpartial_\htau \P + 2\v^k \hpartial_k \P,\nn\\
\hat{g}^{(4)}_{ij} = &\,(1-{\r}) \Big[\(\v_i \v_j-\hs_{ij}\) (\v^2+2\P)%-\frac{1}{r_c}\s_{ij}(v^2+2P)
 +{2}\v_{(i}\hp_{j)}\P -\v_{(i}\hp_{j)}\v^2-(\r+1)\v_{(i}\hp^2 \v_{j)}\nonumber\\
&\qquad\quad -{2}\hs_{ik}\hs^k_{~j}-4\hs_{k(i}\homega^k_{~j)}+({\r-5})\homega_{ik}\homega^k_{~j} -4\hp_i\hp_j
\P +\frac{\r+5}{2}\hp^2\hs_{ij}\Big]. \label{hg42}
\end{align}
The incompressible Navier-Stokes equations (\ref{ins}) change into
\be
\label{hins}
\hp_i \v^i =\htheta,
\quad\p_\htau \v_i +\v^j \hp_j \v_i -\hp^2 \v_i +\hp_i \P = \hf_i,
\ee
where
\begin{align}
\htheta=&\, \l^2\[ -\v^i \hp^2 \v_i +{2}\hs_{ij}\hs^{ij}+\v^i \hp_i \P\] + O(\l^4),\label{htheta}\\
\hf_i=&\,\l^2\[-\frac{3}{2}\hp^4 \v_i+2\v^k \hp^2 \hp_k \v_i+4\hs_{ik}\hp_l \hs^{kl}-{10}\homega_{ik}\hp_l
\hs^{kl}-{3}\hp_i (\hs_{kl}\hs^{kl})- \frac{5}{2}\hp_i (\homega_{kl}\homega^{lk}) \right.\nn \\
&\left.~~~~+4 \hs^{kl} \hp_k \hs_{li}-2\v^k \hp_k \hp_i \P-2 (\hp^k \v_i)\hp_k \P-(\P+\frac{1}{2}\v^2) \hp^2
\v_i  - (\hp_k \hs_{il})\v^k \v^l \right.\nn \\
&\left.~~~~+(\hp_k \homega_{il}) \v^k \v^l+4(\hp_k \v_i )\homega^{kl}\v_l + (\P+\v^2) \hp_i \P - \v_i\p_\htau
\P \]  + O(\l^4)\, .\label{hfi}
\end{align}
With these constraints the metric (\ref{metricnear1}) solves the vacuum Einstein equations (\ref{Einstein}) up
to order $\l^0$ consistently. To solve the next non-trivial order that at $\l^2$, especially the $\htau\htau$
and $\htau i$ components, the terms ${\hat{g}}^{(4)}_{\htau i}$ and ${\hat{g}}^{(4)}_{\htau\htau}$ are needed.
We do not intend to find their explicit expressions here, as it is found that at order $\l^2$, they do not
contribute to the Petrov type \I  equation in (\ref{petrovij}).

\section{The dual Fluid}
\label{AB}
To discuss the fluid dual to vacuum Einstein gravity,
%that the energy density vanishes for equilibrium configurations
the theory of relativistic hydrodynamics up to second order in fluid gradients was presented
in \cite{Compere:2011dx,Chirco:2011ex,Compere:2012mt}. Choosing the Landau frame of the relativistic fluid with
velocity $u^a$ so that
its stress tensor is written as %in %the general form
\be
\label{hydroT}
 T_{ab} = \erho\,u_a u_b + \pp h_{ab} + \Pi^\perp_{ab}, \qquad u^a\Pi^\perp_{ab}=0,
\ee
where $\erho$ and $\pp$ represent the energy density and pressure of the fluid in the local rest frame. The
induced metric $h_{ab} = \g_{ab}+u_au_b$, with $\g_{ab}$ the intrinsically flat metric and $\g_{ab}u^au^b=-1$.
The dissipative corrections can be written down through taking the isotropic gauge so that $\Pi^\perp_{ab}$
does not contain terms proportional to $h_{ab}$. Up to second order in gradients,
\begin{align}\label{PiT}
\Pi^\perp_{ab} &= -2\eta \K_{ab} + \pp^{-1}\[c_1 \K_{ca}\K^c_{~b} + c_2 \K_{c(a}\Omega^c_{~b)} + c_3
\Omega_{ac}\Omega^c_{~b} +c_4 h_a^ch_b^d\D_c\D_d \ln \pp \right.\nn\\
& \left.\quad  + c_5 \K_{ab}\,D\ln \pp + c_6 D^\perp_a \ln \pp \,D^\perp_b\ln \pp\],
\end{align}
where $D^\perp_a\equiv h_a^{\,b}\p_b$, $D\equiv u^a\p_a$ have been defined,
$\eta$ is the relativistic kinematic shear viscosity, and $c_1,...,c_6$ are the corresponding transport
coefficients at the second order. The equations of motion $\D^b T_{ab}$ at the lowest order have been
considered in writing down the above form, and the relativistic shear and vorticity are defined as
\be
\label{KOdef}
 \K_{ab} = h_a^c h_b^d\D_{(c}u_{d)}, \qquad  \Omega_{ab} = h_a^c h_b^d\D_{[c}u_{d]}.
\ee
The energy density which vanishes for equilibrium configurations can also be expanded as
\begin{align}\label{PeT}
\erho = \zeta' D\ln p + \pp^{-1}\[ d_1 \K_{ab}\K^{ab} + d_2 \Omega_{ab}\Omega^{ab} + d_3(D \ln \pp )^2  +d_4 DD
\ln \pp +d_5 (D_\perp \ln \pp )^2 \],
\end{align}
where $\zeta'$ is an alternative first order transport coefficient which is similar to the bulk viscosity that
measures variations of the energy density, and $d_1,...,d_5$ are the corresponding second order transport
coefficients. However, these six coefficients are not independent~\cite{Compere:2012mt}, if we consider the
equation of state of this special fluid dual to vacuum Einstein gravity that $T^2-pT_{ab}T^{ab}=0$, which comes
from the Hamiltonian constraint (\ref{Ham}). Taking account of the expansions \eqref{hydroT} and (\ref{PiT}),
one finds the energy density $\erho$ can be expressed as
\be
\label{dresult1}
\erho = -{2\eta^2}{\pp^{-1}}\K_{ab}\K^{ab}+O(\D^3).
\ee
Comparing (\ref{PeT}) with (\ref{dresult1}), one can read off
\bea
\zeta' = 0 ,\quad d_1 = -{2\eta^2}, \quad d_2=d_3=d_4=d_5 = 0,
\eea
Thus, in this paper we only consider the independent transport coefficients in (\ref{PiT}).

\subsection{Non-relativistic hydrodynamic expansion}
\label{B1}
With the pressure $\pp = r_c^{-1/2} + r_c^{-3/2}P$, the full fluid stress tensor \eqref{hydroT} can be
expanded up to order $\ep^4$ through the non-relativistic hydrodynamical expansion (\ref{orders}) as
\begin{align}
{T^\tau}_i=&+r_c^{-3/2}v_i+r_c^{-5/2}\[v_i(v^2+P)-2\eta r_c\s_{ij}v^j\]+O(\eps^5),\label{Tti1}\\[1ex]
{T^\tau}_\tau=& -r_c^{-3/2}v^2 -r_c^{-5/2}\left[v^2(v^2+P)-2\eta r_c \s_{ij}v^iv^j-{2\eta^2
r_c^2}\s_{ij}\s^{ij}\right]+O(\eps^6),\\[1ex]
{T}_{\,ij}=&+ r_c^{-1/2}\,\delta_{ij}+ r_c^{-3/2}\[P\delta_{ij}+v_iv_j-2\eta r_c \D_{(i} v_{j)}\]\nn\\[1ex]
&+ r_c^{-5/2}\[ v_iv_j (v^2+P)-{\eta r_c}\s_{ij}v^2+2\eta r_c v_{(i}\p_{j)}P -\eta r_c
v_{(i}\p_{j)}v^2-2\eta^2r_c^2 v_{(i}\p^2 v_{j)} \right.\label{Tijc1}\nn\\[1ex]
&\left.+ \,{c_1\,r_c^{2}}\s_{ik}{\s^k}_{j}+c_2\,r_c^{2}{\s}_{k(i}{\omega^{k}_{~j)}}+c_3\,r_c^{2}
\omega_{ik}{\omega^k}_{j}+c_4\, r_c^{2}\p_i\p_j P\]+O(\eps^6),\\[1ex]
{T}\ =&\,{T^\tau}_\tau+{T^i}_{i}= p\,r_c^{-1/2}+p\,r_c^{-3/2}P+O(\eps^6),\label{Ttc1}
\end{align}
where the equations of motion $\p^bT_{ab}=0$ at lower orders have been employed.

\subsection{Alternate presentation}
\label{B2}
With the coordinates in (\ref{hatx}), considering the re-scaled stress tensor
\be\hT_{ab}\d \x^a \d \x^b={r_c^{-1} \e\, } T_{ab}\d x^a\d x^b,\quad \lambda^2\equiv{r^{-1}_c}{
\e^2}\label{nearBY},\ee
one finds the stress tensor (\ref{Tti1})-(\ref{Ttc1}) can be transformed into
\begin{align}
{\hT^\htau}_{~i}=&+\lambda v_i+\lambda^3\[\v_i(\v^2+\P)-2 \heta \hs_{ij}\v^j \]+O(\l^5),\label{ahTti}\\[1ex]
{\hT^\htau}_{~\htau}=& -\l v^2
-\l^3\left[\v^2(\v^2+\P)-2\heta\hs_{ij}\v^i\v^j-2\heta^2\hs_{ij}\hs^{ij}\right]+O(\l^5),\label{ahTtt} \\[1ex]
{\hT}_{\,ij}=&+ \l^{-1}\,\delta_{ij}+ \l\[\P\delta_{ij}+\v_i\v_j-2 \heta \hs_{ij}\]\nn\\[1ex]
&+ \l^{3}\[ \v_i\v_j (\v^2+\P)-\heta\hs_{ij}\v^2+2 \heta\v_{(i}\hp_{j)}\P -\heta
\v_{(i}\hp_{j)}\v^2-2\heta^2\v_{(i}\hp^2 \v_{j)} \right.\nn\\[1ex]
&\left.+\hc_1\,\hs_{ik}{\hs^k}_{~j}+\hc_2\, {\hs}_{k(i}\homega^k_{~j)}+\hc_3\,
\homega_{ik}{\homega^k}_{~j}+\hc_4 \, \hp_i\hp_j \P\]+O(\l^5),\label{ahTij}\\[1ex]
{\hT}\ =&\,{\hT^\htau}_{~\htau}+{\hT^i}_{~i}= \l^{-1}p+\l\,p\,\P+O(\l^5).\label{ahTt}
\end{align}
This is also used to compare with the Brown-York stress tensor dual to the metric (\ref{metricnear1}), which is
mathematically equivalent to the metric with the near horizon expansion~\cite{Bredberg:2011jq}.

\end{document}